\documentclass{aa}
\usepackage{psfig,aabib99}

\begin{document}

   \thesaurus{06 	    
              (08.06.2;     
	       08.19.4;     
	       08.05.3)}    
 
   \headnote{Letter to the Editor} \title{Constraints on mass ejection
             in black hole formation derived from black hole {X}-ray binaries}

   \author{G. Nelemans\inst{}, T. M. Tauris \inst{} and E.P.J. van den Heuvel\inst{}}

   \offprints{gijsn@astro.uva.nl}

   \institute{Astronomical Institute, ``Anton Pannekoek'', University
              of Amsterdam, Kruislaan 403, NL-1098 SJ Amsterdam, The
              Netherlands }

   \date{Received 31 August 1999 / Accepted 3 November 1999}

   \maketitle
   \markboth{Nelemans, Tauris \& van~den~Heuvel: Mass ejection in black hole formation}{}

\begin{abstract} Both the recently observed high runaway velocities of Cyg
{X}-1 ($\sim$ 50 km~s$^{-1}$) and {X}-ray Nova Sco 1994 ($\geq$ 100
km~s$^{-1}$) and the relatively low radial velocities of the black
hole X-ray binaries with low mass donor stars, can be explained by
symmetric mass ejection in the supernovae (SNe) which formed the black
holes in these systems.

Assuming symmetric mass ejection in black hole formation, we estimate
the amount of mass that must have been e\-jec\-ted in the stellar core
collapse in order to explain the velocities of the above {X}-ray
binaries.  We find that at least $2.6\,M_{\odot}$ and $4.1\,M_{\odot}$
must have been ejected in the formation of Cyg {X}-1 and Nova Sco,
respectively. A similar mass loss fraction ($f = 0.35$ ) for the black
hole binaries with low mass donors, gives low velocities, in
agreement with the observations.

We conclude that the black holes in X-ray binaries are all consistent
with being formed in a successful SN in which mass is ejected. A
possible kick at the formation of the black hole is not needed to
explain their space velocities.

\end{abstract}

The light curve and spectrum of the abnormally luminous type Ic SN
1998bw (Galama et al. \cite*{gvp+98}) suggest that in this event
a black hole was formed (Iwamoto et al. \cite*{imn+98}). The
observations imply that a considerable fraction of the mass of the
progenitor (a massive C/O core) was ejected in the explosion (Iwamoto
et al. \cite*{imn+98}). Similarly, the observed overabundance of
the elements O, Mg, Si and S in the atmosphere of the companion of
Nova Sco 1994, indicates considerable mass ejection in the formation
of this black hole \cite{irb+99}.
 
From a study of the
$z$-distribution of the population of black hole X-ray binaries with
low mass donors, White \&\ van~Paradijs \cite*{wp96} conclude that the
velocity dispersion of these X-ray binaries is of the order of 40
km~s$^{-1}$. Since the velocity dispersion of the progenitor systems is
expected to be around 17 km~s$^{-1}$, they estimate the extra velocity that
is given to the system in the formation of the black hole to be 20~--
~40 km~s$^{-1}$. This requires substantial mass ejection in the formation of a
black hole if no asymmetric kicks are involved. 

Recent determinations of the space velocity of
Cyg {X}-1 \cite{kcb99} and the radial velocity of
Nova Sco \cite{bom+95} demonstrate that these black hole 
binaries have significantly higher runaway velocities than the
black hole X-ray binaries with low mass donors.

In
Table~\ref{tab:properties} we have listed the relevant properties of
the galactic black hole binaries for which one, or more, of its
velocity components have been measured.
 
\begin{table}
\caption[]{Properties of black hole X-ray binaries. The velocity of
Cyg X-1 is its space velocity, all other velocities are radial
velocities, so are lower limits}
\label{tab:properties}
\begin{tabular}{lrrrrr}
Source 		&  $M ({\rm M}_{\sun})$ & $m ({\rm M}_{\sun})$ & $P$ (d) & $v$ (km~s$^{-1}$) \\ \hline
Nova Sco 1994	&  6.29-7.60   & 1.6 - 3.1   & 2.62    & 106$\pm$19 \\
Cyg X-1		&  3.9-15.2    & 11.7-19.2   & 5.6     & 49$\pm$14  \\ \hline
V 404 Cyg	&  6-12.5      & 0.6         & 6.5     & 8.5$\pm$2.2 \\
A 0620-00	&  3.3-4.24    & 0.15-0.38   & 0.32    & 15$\pm$5   \\
Nova Muscae	&$>$4.45$\pm$0.46& 0.7       & 0.43    & 26$\pm$5   \\
Nova Oph 1977	&  5-7         & 0.7         & -       & 38$\pm$20  \\
GRO J0422+32	&  3.25-3.9    & 0.39        & 0.21    & 11$\pm$8   \\
GS2000+25	&  6.04-13.9   & 0.26-0.59   & 0.35    & 18.9$\pm$4.2\\
		&              &             &         &            
\end{tabular}\\
Masses and periods from Ergma \&\ van~den~Heuvel~\cite*{eh98a} and
references therein, velocities from Brandt~et~al.~\cite*{bps95} and
references therein, except for GRO J0422+32 \cite{hch+99} and
GS2000+25 \cite{hhf96} which are heliocentric $\gamma$-velocities. For
Nova Sco 1994 we changed the velocity according to the new
$\gamma$-velocity of 142 km~s$^{-1}$ \cite{shc+99}. Space velocity for Cyg
X-1 is from Kaper~et~al. \cite*{kcb99}.
\end{table}

\section{Origin of the black hole binary runaway velocities}\label{origin}

There are two effects to accelerate a binary system by a
supernova explosion. The first is caused by the ejection of material
from the binary (\nocite{bla61}Blaauw 1961). The centre of mass of
the ejected matter will continue to move with the orbital velocity of
the black hole progenitor. To conserve momentum, the binary will move
in the opposite direction. The second one is an additional velocity
kick, which is produced by asymmetries in the supernova explosion
itself and for which there is  strong evidence in the case of the
formation of a neutron star (e.g. Lyne \&\ Lorimer~\cite*{ll94},
Hartman~\cite*{har97b}).

The current status quo of supernova simulations is that in order
to get a successful supernova, in which the shock is reversed and
matter is ejected, one needs to form a neutron star \cite{bw85}. If
the supernova is not so energetic, there may be considerable fall
back, turning the neutron star into a black hole (e.g. Colgate
\cite*{col71}; Woosley \&\ Weaver \cite*{ww95}). Formation of a black
hole without an intermediate neutron star would then not result in mass
ejection. 
However, if other mechanisms than neutrino
heating will be found to reverse the supernova shock (e.g rotation),
this conjecture of both mechanisms may be broken.

Brandt~et~al.~\cite*{bps95} have listed a number of scenarios for
reproducing the high radial velocity measured in Nova Sco. They
show that though mass ejection alone can explain the velocity of
Nova Sco 1994, the allowed range of initial masses is very small. They
therefore conclude that Nova Sco is formed in a delayed black hole
formation, in which the kick, which is imparted to the initial neutron
star, is responsible for a considerable fraction of the present system
velocity. The black holes in the other binaries would then be formed
by a direct collapse without mass ejection and kicks. The velocity
dispersion found by White \&\ van Paradijs \cite*{wp96} can be
explained by scattering at molecular clouds and density waves, since
these binaries could be an old population (Podsiadlowski, private
communication; see also Brandt~et~al. \cite*{bps95}).

With the new discovery of the relatively high velocity of Cyg X-1, we think 
the above is unlikely, because now the two systems with
highest mass companions must have formed through a delayed black hole
formation, while the systems with low mass companions form in a direct
collapse. 
This would mean that the success of the SN in which the black hole is
formed is related to the nature of its binary companion, for which we
see no reason

Tutukov \&\ Cherepahshchuk \cite*{tc97} discuss the system velocities
of the X-ray binaries containing black holes and conclude that all
velocities can be explained with mass ejection alone. However, they
only consider the maximum velocity that can be obtained with the
observed limits on the masses of both stars, assuming the shortest
possible period at the moment of the SN and the maximum amount of mass
that can be ejected without disrupting the binary.
In that case, the pre-SN mass ratio is not independent of the final
(observed) mass ratio and it would be better to use the current
(observed) mass ratio, with which their equation (7) would become
\begin{equation}\label{eq:tutukov}
v_{\rm max} = 192 \left(\frac{q_{\rm obs}}{1 + q_{\rm
obs}}\right)^{0.72} \qquad {\rm km \; s^{-1}}
\end{equation} 
See also the discussion in section~\ref{conclusion}.

We now investigate the effect of the mass ejection in more
detail, assuming possible kicks are (relatively) unimportant.

\section{Runaway velocities from symmetric SNe}

\begin{figure*}[t]
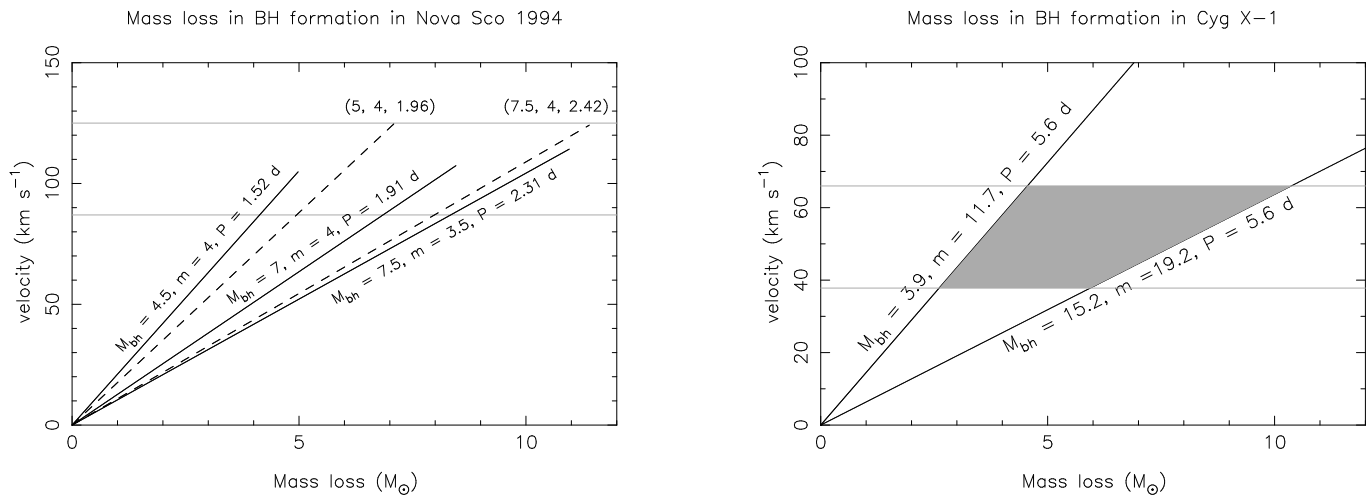

\begin{minipage}{0.45\textwidth}
\psfig{figure=Bh302.f1.ps,width=\textwidth,angle=-90}
\end{minipage}
\hspace*{0.095\textwidth}
\begin{minipage}{0.45\textwidth}
\psfig{figure=Bh302.f2.ps,width=\textwidth,angle=-90}
\end{minipage}
\caption{Limits on the amount of mass ejected in the SN explosion
that is required to explain the measured velocities. Left: Nova Sco
1994 with three possibilities of the binary parameters at the onset of
the X-ray phase (solid lines; see text) and the two possibilities in
the case 0.5 ${\rm M}_{\sun}$ has been lost draining angular momentum (dashed
lines).  Right: Cygnus X-1 with two different solutions for the
companion mass.}
\label{fig:mass_lost}
\end{figure*}

Consider a circular pre-SN orbit consisting of a helium star
with mass $M_{\rm He}$ (the progenitor of the black hole)
and a companion star with mass $m$. Assume that the helium star
explodes in a symmetric SN during which an amount of mass,
$\Delta M$ is ejected instantaneously and decouples gravitationally
from the system.
If $\Delta M = M_{\rm He} - M_{\rm BH} < 0.5\,(M_{\rm He}+m)$
the binary will remain bound. The post-SN eccentricity, period and
orbital separation are given by Bhattacharya \&\ van~den~Heuvel~\cite*{bh91}
\begin{equation}\label{eq:eccentricity}
e_{\rm postSN} = \frac{\Delta M}{M_{\rm BH} + m} = \frac{ 1 - \mu}{\mu}
\end{equation}
\begin{equation}
P_{\rm postSN} = P_{\rm i} \; \frac{\mu}{(2\mu - 1)^{3/2}}
\end{equation}
where we define
\begin{equation}
\mu = \frac{M_{\rm BH} + m}{M_{\rm He} + m} = \frac{M_{\rm He} + m - \Delta M}{M_{\rm He} + m}
\end{equation}
and subscripts $i$ denote the pre-SN system. Since the observed black
hole binaries all have short orbital periods ($<\,7$ days) tidal
forces act to re-circularize the post-SN orbit. The parameters of the
re-circularized orbit are given by
\begin{equation}
P_{\rm re-circ} = P_{\rm postSN} (1 - e_{\rm postSN}^2)^{3/2} = P_{\rm i} / \mu^2
\end{equation}
And similarly $a_{\rm re-circ} = a_{\rm i} / \mu$.
Here we have ignored the effects of the impact of the ejected shell
on the companion star and assume 
there is no mass loss or transfer
during the re-cir\-cu\-la\-riza\-tion phase.
From conservation of momentum one finds an expression for the
resulting runaway velocity (recoil) of the system
\begin{equation}\label{eq:momentum}
  v_{\rm sys} = \frac{\Delta M \, v_{\rm He}}{M_{\rm BH}+m}
\end{equation}
where $v_{\rm He}$ is the pre-SN orbital velocity of the exploding
helium star in a centre-of-mass reference frame. Together with Keplers
third law we find
\begin{equation}\label{eq:v_sys}
  v_{\rm sys} = (G\,2\pi)^{1/3}\,\Delta M\,m\,
                P_{\rm re-circ}^{-1/3}\,(M_{\rm BH}+m)^{-5/3}
\end{equation}
For convenience this equation can be expressed as
\begin{displaymath}
v_{\rm sys}
 = 213 
 \! \left(\!\frac{\Delta M}{M_{\odot}}\!\right) \!\!\left(\!\frac{m}{M_{\odot}} 
\!\right) \!\!\left(\! \frac{P_{\rm re-circ}}{\mbox{day}} \!\right)^{\!\!\!-\frac{1}{3}} \!\!\left(
\frac{\!M_{\rm BH}\!+\!m}{M_{\odot}} \!\right)^{\!\!\!-\frac{5}{3}}\! \mbox{km s$^{-1}$}
\end{displaymath}

If we know the masses of the stellar components and the orbital period
after the re-circularization ($M_{\rm BH}, m, P_{\rm re-circ}$)
we can calculate $\Delta M$ from the observed runaway velocity,
$v_{\rm sys}$. However, we observe mass-transferring binaries which
might have evolved due to loss of angular momentum by gravitational
radiation or magnetic braking before the mass
transfer started and/or might have transferred already a significant
amount of mass from the donor to the black hole. Before applying
equation~(\ref{eq:v_sys}) to the observed systems we have to correct
for these effects.

Also, one has to check whether the binary before the SN would be
detached, i.e. that both stars do not fill their Roche lobes at the
moment the SN explodes.

\section{Results}\label{results}

\subsection{Nova Sco 1994}
Shahbaz~et~al., \cite*{shc+99} have recently determined the present
stellar masses in Nova Sco 1994 (GRO J1655-40). They find $M_{\rm
BH}=5.5-7.9\,M_{\odot}$ and $m=1.7-3.3\,M_{\odot}$.  The mass transfer
in Nova Sco 1994 may already have been going on for a long period of
time. From the luminosity and effective temperature of the donor star
in this system one finds, using stellar evolution tracks, that the
donor can not have started out with a mass larger than
$4.0\,M_{\odot}$ at the onset of the {X}-ray phase \cite{hha+98}. As
an example of combinations of present masses we use ($M_{\rm BH}, m$)
= (6, 2.5) and (7.75, 3.25). Assuming conservative mass transfer
($P_{\rm re-circ}/P_{\rm obs} = [(M_{\rm BH, obs} m_{\rm obs})/$ $(M_{\rm
BH} m)]^3$) some possibilities for the system configuration at the
onset of mass transfer are the following combinations of $(M_{\rm BH},
m, P_{\rm re-circ})$: (4.5, 4.0, 1.52), (7.0, 4.0, 1.91) and (7.5,
3.5, 2.31). With these values and Eq.~(\ref{eq:v_sys}) we find that
the present runaway velocity of 106 km~s$^{-1}$ is obtained for
$\Delta M = 5.0\,, 8.4$ and $10.2 M_{\odot}$ respectively. This is
shown in the left panel of Fig.~\ref{fig:mass_lost} (solid lines).
Note that all these lines are terminated at the amount of mass
ejection that would result in a pre-SN orbit in which the companion
would fill its Roche lobe. The minimum amount of mass that must be
lost is 4.1 ${\rm M}_{\sun}$ in the case of a black hole of 4.5 ${\rm M}_{\sun}$, given
$v_{\rm sys} > 87$ km s$^{-1}$.

If we relax the assumption of conservative
mass transfer (as is suggested by the observation of jets from Nova 
Sco), and assuming the lost material drags along three times
the specific angular momentum (Pols \&\ Marinus \cite*{pm94}), we
calculated the orbits for the first two cases, assuming 0.5
$M_{\odot}$ was lost from the system. The resulting system parameters 
at the onset of the mass transfer then become (5, 4, 1.96) and (7.5,
4, 2.42) for the first two examples. These curves are plotted as
dashed lines in Fig.~\ref{fig:mass_lost}. In this case at least  5
and 8 ${\rm M}_{\sun}$ are lost, respectively.


\subsection{Cygnus X-1}
For Cygnus X-1 the presently best estimate of the masses of the
stellar components is $M_{\rm BH}=10.1\,M_{\odot}$ and $m =
17.8\,M_{\odot}$. Extremes of the allowed masses are given by
$(M_{\rm BH}, m)$ = (3.9, 11.7) and (15.2, 19.2) respectively
\cite{hkg+95}.
We assume no orbital evolution since the beginning of the mass
transfer phase, because Roche lobe overflow
can not have started long ago since the expected mass transfer rates
then would be much higher.  We use the values of the masses as given
above and the present day orbital period of 5.6 days. We also neglect
the small eccentricity that the orbit still has. The right hand panel
in Fig.~\ref{fig:mass_lost} shows the resulting allowed range of
mass ejected in the formation of the black hole. For a present black
hole mass of 3.9 ${\rm M}_{\sun}$ at least 2.6 ${\rm M}_{\sun}$ must have been ejected to
produce the observed space velocity. For a black hole of 15.2 ${\rm M}_{\sun}$ at
least 6 ${\rm M}_{\sun}$ must have been ejected.

\subsection{The remaining black hole {X}-ray transients}

\begin{figure}[t]
\psfig{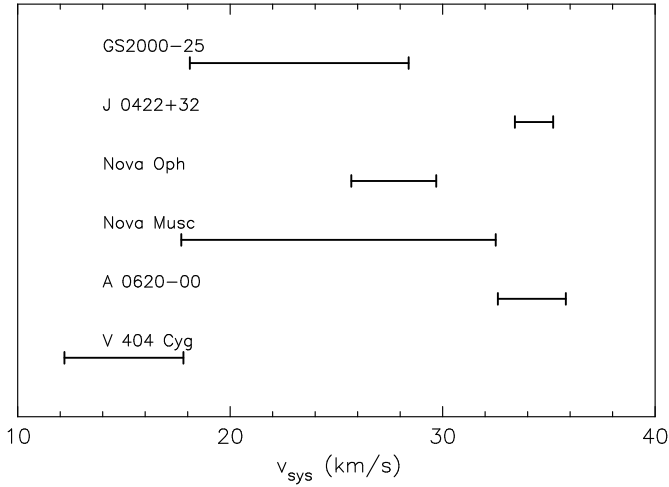}
\caption{Our estimated 3-D recoil velocity for the black hole X-ray
binaries with low mass donors, for a supernova mass loss fraction $f
\equiv \Delta M/M_{\rm He} = 0.35$. The limits represent the uncertainty in the
black hole mass as given in Table~\ref{tab:properties}}
\label{fig:v_ml}
\end{figure}

Table~\ref{tab:properties} shows that all the black hole X-ray
binaries with low mass donors have low velocities. As
derived by White \&\ van~Pa\-ra\-dijs \cite*{wp96}, the expected
additional velocity component of these X-ray binaries is of the order
of 20 -- 40 km~s$^{-1}$. 
In Cyg~X-1 and Nova~Sco at least 28 and 48\% of the mass of the
progenitor must have been ejected in the SN.
Therefore we computed the velocities for these systems assuming a
constant fraction of 35\% of the helium star mass to be ejected and
show the obtained range in velocities given the range in black hole
masses in Fig.~\ref{fig:v_ml}.
The last five systems
are all expected to have evolved during mass transfer to smaller
periods, and all seem to be compatible with an initial systems close
to ($m$, $P$) = (1 ${\rm M}_{\sun}$, 0.74 d), cf. Ergma
\&\ Fedorova \cite*{ef98}. The systems shrink due to magnetic braking,
so we assume $P_{\rm re-circ}$ = 1 d (see e.g. Kalogera
\cite*{kal99}). \pagebreak[0]For V 404 Cyg we assumed an
re-circularized period of 4 days and an donor mass of
1 ${\rm M}_{\sun}$.

\section{Discussion and conclusions}\label{conclusion}

In this Letter we show that both the high observed velocities of Cyg
X-1 and Nova Sco, and the low velocities of the black hole X-ray
binaries with low mass donors can be explained by ejection of $\ga
$30\% of the mass of the exploding helium star in the SN that formed
the black hole. This removes the need to invoke a large kick for Nova
Sco (and Cyg X-1) and at the same time a small or no kick for the
remaining systems \cite{bps95}, which seems highly unlikely to us.

The radial velocity of Nova Sco can only be explained by large
mass-loss fractions ($\ga 50\%$) and the assumption that the
mass transfer has already started some time ago. If this mass transfer 
was non-conservative (consistent with observed jets), the velocity can
be explained more easily. This may also be needed if Nova Sco also has 
a transverse velocity component. 

Tutukov \&\ Cherepahschuk \cite*{tc97} state that the high velocity of
Nova Sco could be obtained by having the pre-SN mass ratio above 0.24,
i.e. $M_{\rm He} \leq 9.6$ (they use 2.3 and 4 ${\rm M}_{\sun}$ for the current
masses). However, this is not in agreement with the
assumption in their equation, that the maximal amount of mass is
lost. Using our modification (Eq. (\ref{eq:tutukov})) to their
equation, we find indeed that for their masses it is impossible to
obtain a velocity higher than 93 km~s $^{-1}$, in agreement with our
findings that the post-SN orbit must be different from the current
orbit.

The fact that black holes in X-ray binaries
may have lost several tens of percents of their progenitor mass in the
SN, makes it necessary that some of their progenitor (helium) stars
must have had masses above 10 ${\rm M}_{\sun}$, which is in clear disagreement
with the suggestion from some stellar evolution models that all
Wolf-Rayet stars have a mass $\la 3.5
{\rm M}_{\sun}$ at the moment they explode in a supernova \cite{wlw95}.

Finally, it should be noticed that the conclusion that black holes
eject a substantial amount of material during their formation has
consequences for the orbital period distribution of close
black hole pairs, which are expected to be prime sources for ground
based gravitational wave detectors. Mass ejection will widen the
orbit, which happens twice during the formation of a black hole pair,
possibly preventing black holes to form in a close orbit at all. Only
kicks from an asymmetry in the SN could then form close pairs. But as
shown above, there is not much evidence for kicks and the
magnitude of any kicks is severely limited to $<$ 40 km~s$^{-1}$ by the black
hole X-ray binaries with low mass donors, unless the black holes in
these system formed in a direct collapse.
\begin{acknowledgements}
We thank Philipp Podsiadlowski and the referee for comments that
improved this article and Lex Kaper who made us aware of the
proper motion of Cyg X-1.  This work was supported by NWO Spinoza
grant 08-0 to E.P.J. van den Heuvel.
\end{acknowledgements}

\bibliography{journals,binaries}
\bibliographystyle{aabib99}

\end{document}